\documentclass[a4paper]{amsproc}
\usepackage{amssymb}
\usepackage{amscd}

\usepackage[all,cmtip]{xy}

\usepackage{amsfonts}
\usepackage{cite}
\usepackage{amsmath}
\usepackage{epsfig}
\usepackage{graphicx, subfigure}
\usepackage{color, graphics}
\usepackage{pdfsync}
\usepackage{hyperref}

\allowdisplaybreaks

\theoremstyle{plain}
 \newtheorem{thm}{Theorem}
 \newtheorem{thmA}{Main result}
 \newtheorem{prop}{Proposition}

\theoremstyle{definition}
 
 \newtheorem{remark}{Remark}

\numberwithin{equation}{section}

\newcommand{\R}{\mathbb{ R}}

\DeclareMathOperator{\pr}{pr}

\DeclareMathOperator{\diag}{diag}

\title[Integrable spherical and planar ball bearings] {Spherical  and Planar Ball Bearings -- a Study of Integrable Cases }

\author[V. Dragovi\'c, B. Gaji\'c, B. Jovanovi\'c]{\bfseries Vladimir Dragovi\'c, Borislav Gaji\'c, Bo\v zidar Jovanovi\'c}

\address{
Department of Mathematical Sciences  \\
The University of Texas at Dallas   \\
Richardson, TX\\
USA\\
Mathematical Institute\\
Serbian Academy of Sciences and Arts\\
Belgrade\\
Serbia}
\email{Vladimir.Dragovic@utdallas.edu}

\address{
Mathematical Institute\\
Serbian Academy of Sciences and Arts\\
Belgrade\\
Serbia}
\email{gajab@mi.sanu.ac.rs}

\address{
Mathematical Institute\\
Serbian Academy of Sciences and Arts\\
Belgrade\\
Serbia}
\email{bozaj@mi.sanu.ac.rs}

\subjclass[2010]{37J60, 37J35, 70E40,  70F25}

\keywords{Nonholonimic dynamics; rolling without slipping, invariant measure; integrability}

\begin{document}

\begin{abstract}
We consider the nonholonomic systems of $n$ homogeneous balls $\mathbf B_1,\dots,\mathbf B_n$  with the same radius $r$ that are rolling without slipping about a fixed sphere $\mathbf S_0$ with center $O$ and radius $R$.
In addition, it is assumed that a dynamically nonsymmetric sphere $\mathbf S$ with the center that coincides with the center $O$ of the fixed sphere  $\mathbf S_0$ rolls without slipping in contact to the moving balls $\mathbf B_1,\dots,\mathbf B_n$. The problem is considered in four different configurations,  three of which are new.
We derive the equations of motion and explicitly indicate an invariant measure for these systems.
As the main result, for $n=1$ we found two cases that are integrable in quadratures according to the Euler-Jacobi theorem.
The obtained integrable nonholonomic models are natural extensions of  the well-known Chaplygin ball integrable problems.
Further, we explicitly integrate
 the planar problem consisting of $n$ homogeneous balls of the same radius, but with different masses, that roll without slipping
over a fixed plane $\Sigma_0$ with a plane $\Sigma$  that moves without slipping over these balls.

\end{abstract}

\maketitle

\section{Introduction}

\begin{figure}[ht]\label{Fig1}
\includegraphics[width=115mm]{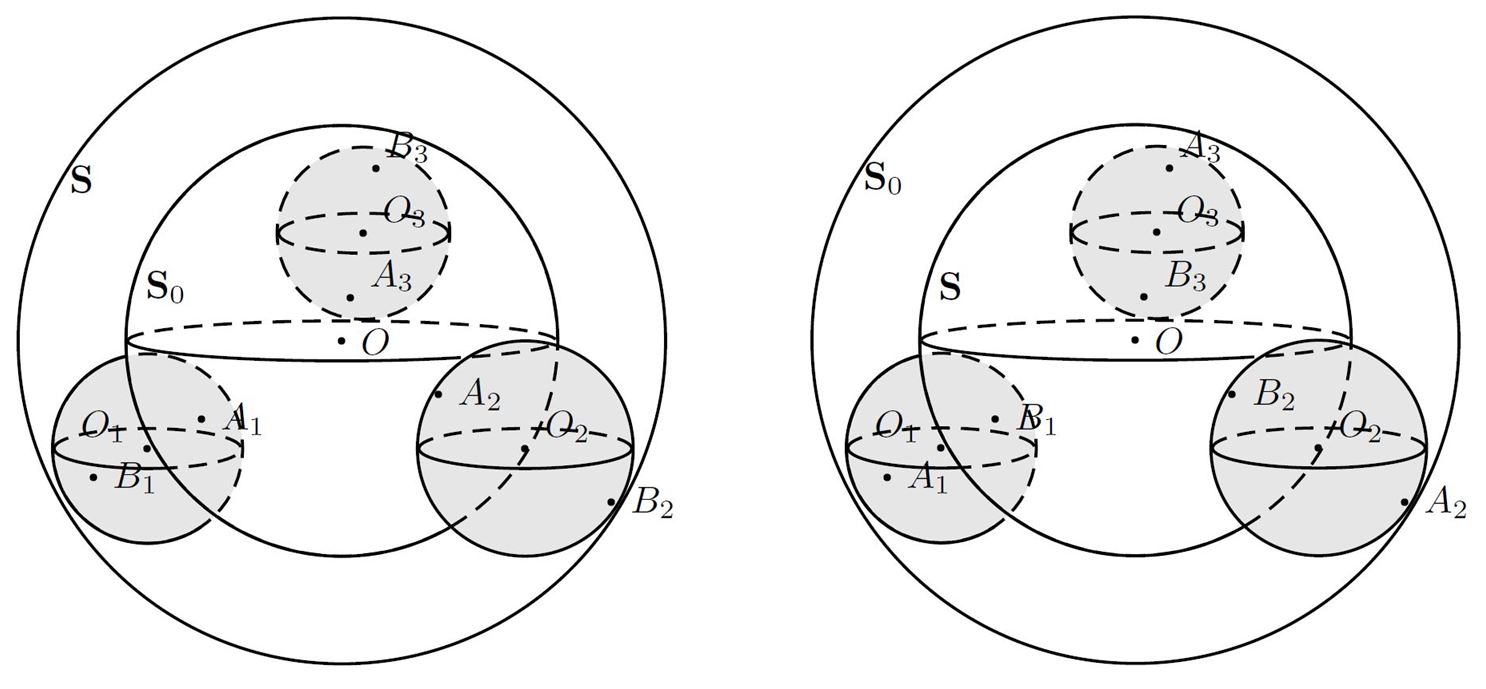}
\caption{Spherical ball bearings
 for $n=3$, case I (left, $\rho=R+2r$)  and case II (right, $\rho=R-2r$)}
\end{figure}

We continue our study the spherical and the planar ball bearing problems, which we introduced in \cite{DGJRCD}.
Here we focus on four different configurations of
the spherical ball bearing problem. In \cite{DGJRCD} we dealt with the first configuration:
$n$ homogeneous balls $\mathbf B_1,\dots,\mathbf B_n$ with centers $O_1,...,O_n$ and the same radius $r$  roll without slipping around a fixed sphere $\mathbf S_0$ with center $O$ and radius $R$. A dynamically nonsymmetric sphere $\mathbf S$ of radius $\rho=R+2r$ with the center that coincides with the center $O$ of the fixed sphere $\mathbf S_0$  rolls without slipping over the moving balls $\mathbf B_1,\dots,\mathbf B_n$
({\bf case I}, Figure \ref{Fig1}).

As the second configuration ({\bf case II}), we consider homogeneous balls of radius $r$ within a fixed sphere $\mathbf S_0$ of radius $R$. The balls support a moving,
dynamically nonsymmetric sphere $\mathbf S$ of radius $\rho=R-2r$ (see Figure \ref{Fig1}).

Proposition \ref{prva} implies that the centers $O_1,...,O_n$ of the balls are in rest in relation to each other. Thus, there are no collisions of the balls $\mathbf B_1,\dots,\mathbf B_n$.
For $n \ge 4$ there are initial positions of the balls $\mathbf B_1,\dots,\mathbf B_n$ that imply the condition that the centre of the moving sphere $\mathbf S$ coincides with the centre $O$ of the fixed sphere $\mathbf S_0$. In order to include all possible initial positions for arbitrary $n$, the condition that $O$ coincides with the centre of the sphere $\mathbf S$ is assumed to be a holonomic constraint.

For $n=1$ we introduce two additional configurations assuming that  $\mathbf B_1$  is not a homogeneous ball but a sphere (spherical shell).
The first one is when the sphere $\mathbf B_1$ is within the moving sphere $\mathbf S$ and the fixed sphere $\mathbf S_0$ is within $\mathbf B_1$ ({\bf case III},
$\rho=2r-R$, $\rho>R$, Figure 2).
The second one is when the sphere $\mathbf B_1$ is within the fixed sphere $\mathbf S_0$ and the moving sphere $\mathbf S_0$ is within $\mathbf B_1$ ({\bf case IV},
$\rho=2r-R$, $\rho<R$, Figure 2).

In Section \ref{sec2} we present the equations of motion of the spherical ball bearing systems for all four configurations.
The kinetic energy and the distribution  are invariant with respect to an appropriate action of the Lie group $SO(3)^{n+1}$,
and the system can be reduced to $\mathcal M=\mathcal D/SO(3)^{n+1}$, where
$\mathcal D\subset TQ$ is the nonholonomic distribution  and $Q=SO(3)^{n+1}\times S^n$ is
the configuration space of the problem.
In addition, by fixing values of first integrals, the system can be  also reduced to a second reduced space $\mathcal N=\R^3\times (S^2)^n$, see Theorem \ref{Glavna}.

The system also has an invariant measure, see Theorem \ref{mera}, Section \ref{sec3}. The proofs of Theorems \ref{Glavna} and \ref{mera} are
similar to the proofs of the corresponding statements given for the configuration I in \cite{DGJRCD}. Thus, they are omitted.

In this paper we consider the integrability of the spherical balls bearing problem in the case of $n=1$. The system can be reduced to $\mathcal N=\R^3\{\vec\Omega\}\times S^2\{\vec\Gamma\}$ and takes the form (see Section \ref{sec3})
\begin{equation}\label{redukovaneI}
\dot{\vec{\mathbf M}}=\vec{\mathbf M}\times\vec\Omega, \qquad \dot{\vec{\Gamma}}=\varepsilon\vec\Gamma\times\vec\Omega,
\end{equation}
where
$\vec{\mathbf M}=\mathbf I\vec{\Omega}+d\vec\Gamma$ and
\[
\mathbf I=\mathbb I+D\mathbf E-D\vec\Gamma\otimes\vec\Gamma,  \qquad \mathbf E=\diag(1,1,1).
\]

Here $\vec\Omega$ is the angular velocity of the sphere $\mathbf S$, $\mathbb I=\diag(A,B,C)$ is its inertial tensor, $\vec\Gamma$ is the unit vector determining the position of the homogeneous ball $\mathbf B_1$ and $\varepsilon$, $d$, $D$ are parameters of the problem that are described in Sections \ref{sec2} and \ref{sec3}.

According to Theorem \ref{mera}, the flow of \eqref{redukovaneI} in variables $\{\vec\Omega,\vec\Gamma\}$ preserves the measure with density $\sqrt{\det(\mathbf I)}$.
Also, it  always has the  first integrals
$F_1=\frac12 \langle \mathbf I\vec\Omega,\vec\Omega\rangle$ and $F_2=\langle \vec{\mathbf M}, \vec{\mathbf  M}\rangle$ (see Proposition \ref{integrali}).
Since $\mathcal N$ is five-dimensional, for the integrability, according to the Euler-Jacobi theorem, one additional first integral is needed.

\begin{figure}[ht]\label{Fig2}
\includegraphics[width=115mm]{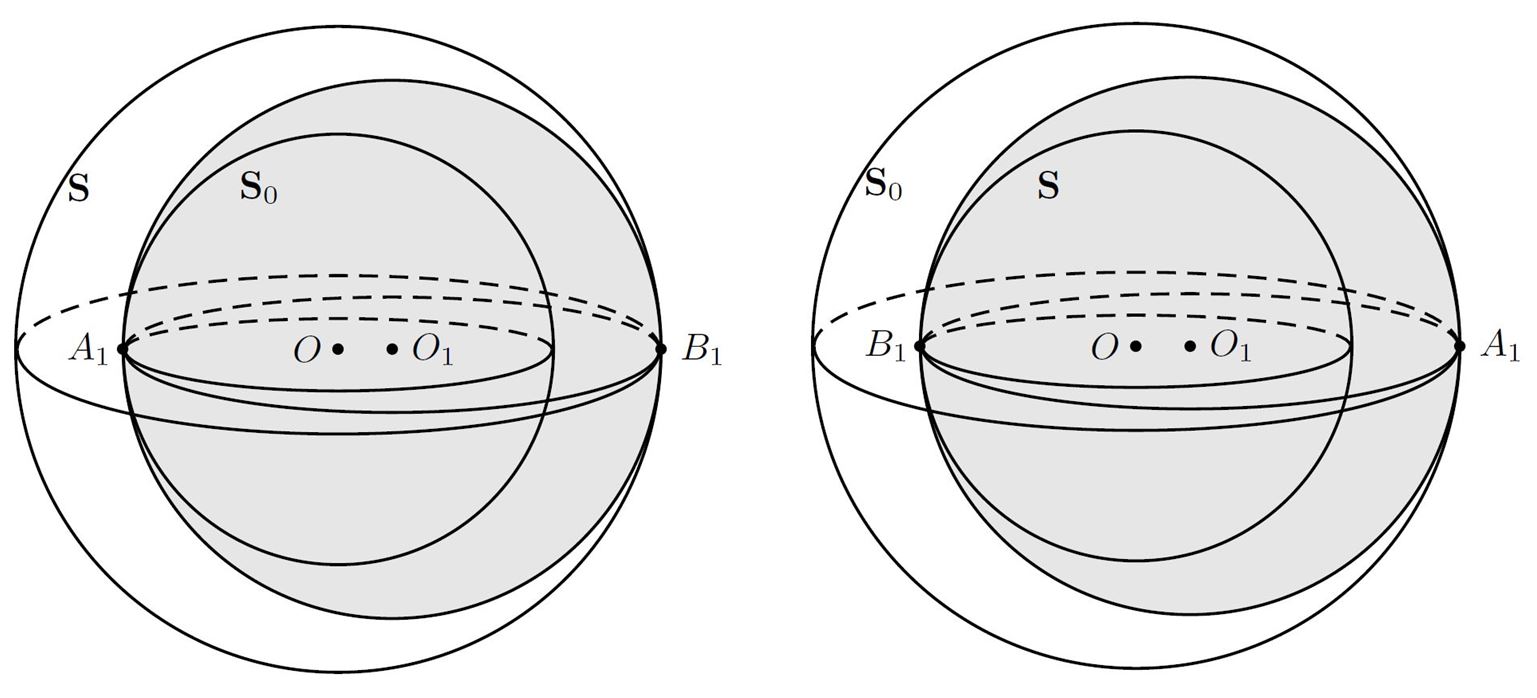}
\caption{Spherical ball bearing, case III (left, $\rho=2r-R$) and case IV (right, $\rho=2r-R$)}
\end{figure}

As the main results of the paper, in Section \ref{sec3} we prove

\begin{thmA}
The spherical ball bearings problem \eqref{redukovaneI} in the configuration III, when $2r=3R$, i.e., $\varepsilon=-1$, is integrable. The third first integral is
\[
F_3=(B+C-A+D) \mathbf M_1\Gamma_1 +(A+C-B+D) \mathbf M_2 \Gamma_2+(A+B-C+D)\mathbf M_3\Gamma_3.
\]
\end{thmA}

\begin{thmA}
The spherical ball bearings problem \eqref{redukovaneI} for $B=C$ is integrable for any $\varepsilon$ in all configurations. Along with $F_1$ and $F_2$, the system
has  two additional nonalgebraic first integrals.
\end{thmA}

 The exact formulae for these additional nonalgebraic first integrals, indicated in the above Theorem, are given in  \eqref{f3pm}.

For $d=0$, the equations \eqref{redukovaneI} coincide with the equations of motion of a Chaplygin ball with the inertia tensor $\mathbb I$ on a sphere ($\varepsilon\ne 1$) and the plane ($\varepsilon=1$) with slightly different definitions of parameters $\varepsilon$ and $D$. Thus, the above results can be seen as natural extensions of well-known
integrable Chaplygin ball problems.

In \cite{DGJRCD} we also considered
the associated planar problem. This is the case I in the notation of the present paper, when the radius of the fixed sphere $\mathbf S_0$ tends to infinity.  We found an
invariant measure and proved the integrability by means of the Euler--Jacobi theorem \cite{DGJRCD}.  Here, in Section 4,
we perform an explicit integration of the reduced problem.

\section{Rolling of a dynamically nonsymmetric sphere over $n$ moving homogeneous balls and a fixed sphere}\label{sec2}

\subsection{Kinematics}
Let  $O\vec{\mathbf e}^0_1,\vec{\mathbf e}^0_2,\vec{\mathbf e}^0_3$, $O\vec{\mathbf e}_1,\vec{\mathbf e}_2,\vec{\mathbf e}_3$,
$O_i\vec{\mathbf e}^i_1,\vec{\mathbf e}^i_2,\vec{\mathbf e}^i_3$
be positively oriented reference frames rigidly attached to the spheres $\mathbf S_0$, $\mathbf S$, and the balls $\mathbf B_i$, $i=1,\dots,n$, respectively.
By $\mathbf g,\mathbf g_i\in SO(3)$ we denote the matrices that map the
moving frames $O\vec{\mathbf e}_1,\vec{\mathbf e}_2,\vec{\mathbf e}_3$ and $O_i\vec{\mathbf e}^i_1,\vec{\mathbf e}^i_2,\vec{\mathbf e}^i_3$
to the fixed frame $O\vec{\mathbf e}^0_1,\vec{\mathbf e}^0_2,\vec{\mathbf e}^0_3$.
Using the standard isomorphism between the Lie algebras $(so(3),[\cdot,\cdot])$ and $(\R^3,\times)$
given by
\begin{equation}\label{izomorfizam}
a_{ij}=-\varepsilon_{ijk}a_k, \qquad i,j,k=1,2,3,
\end{equation}
the skew-symmetric matrices $\omega=\dot{\mathbf g}{\mathbf g}^{-1}$, $\omega_i=\dot{\mathbf g}_i \mathbf g_i^{-1}$
correspond to the angular velocities $\vec{\omega}$, $\vec{\omega}_i$ of the sphere $\mathbf S$ and the $i$-th ball $\mathbf B_i$ in the fixed reference frame
$O\vec{\mathbf e}^0_1,\vec{\mathbf e}^0_2,\vec{\mathbf e}^0_3$ attached to the sphere $\mathbf S_0$.

The matrices $\Omega=\mathbf g^{-1}\dot{\mathbf g}=\mathbf g^{-1}\omega\mathbf g$, $W_i=\mathbf g^{-1}_i\dot{\mathbf g}_i=\mathbf g^{-1}_i\omega_i\mathbf g_i$
correspond to the angular velocities $\vec{\Omega}$, $\vec{W}_i$ of $\mathbf S$ and $\mathbf B_i$ in the frames $O\vec{\mathbf e}_1,\vec{\mathbf e}_2,\vec{\mathbf e}_3$
and  $O_i\vec{\mathbf e}^i_1,\vec{\mathbf e}^i_2,\vec{\mathbf e}^i_3$ attached to the sphere $\mathbf S$ and the balls $\mathbf B_i$, respectively.
We have
$\vec{\omega}=\mathbf g\vec{\Omega}$, $\vec{\omega}_i=\mathbf g_i \vec{W}_i$.

Then the configuration space of the problem is
\[
Q=SO(3)^{n+1}\times (S^2)^{n}\{\mathbf g,\mathbf g_1,\dots,\mathbf g_n,\vec\gamma_1,\dots,\vec\gamma_n\},
\]
where $\vec{\gamma}_i$ is the unit vector
\[
\vec{\gamma}_i=\frac{\overrightarrow{OO_i}}{|\overrightarrow{OO_i}|}
\]
determining the position of the centre of $i$-th ball $\mathbf B_i$, $i=1,\dots,n$.
In the cases I and II, the velocity of the centre
of the $i$-th ball is $\vec{v}_{O_i}=(R\pm r)\dot{\vec\gamma}_i$, while
for the cases III and IV ($n=1$) we have $\vec{v}_{O_1}=\pm(r-R)\dot{\vec\gamma}_1$.\footnote{Through the paper, the sign $\pm$  denotes $+$ for cases I and III, and - for the cases II and IV.}
It follows from  Proposition \ref{prva} that if the initial conditions are chosen such that the distances between $O_i$ and $O_j$ are all greater than $2r$, $1 \le i< j \le n$, then the balls will not have collisions along the course of motion. This is the reason why we do not assume additional one-side constraints
\begin{equation*}\label{oblast}
\vert \vec\gamma_i-\vec\gamma_j \vert \ge \frac{2r}{R\pm r}, \qquad 1\le i<j\le n \qquad \text{(cases I and II, $n\ge 2$)}.
\end{equation*}

Let  $A_1,..., A_n$ and
$B_1, B_2,...,B_n$ be the contact points of the balls $\mathbf B_1,\dots,\mathbf B_n$ with the spheres $\mathbf S_0$ and $\mathbf S$, respectively.
The condition that the rolling of the balls $\mathbf B_1,\dots,\mathbf B_n$ and the sphere $\mathbf S$ are without slipping leads to the nonholonomic constraints:
\[
\vec{v}_{O_i}+\vec{\omega}_i\times\overrightarrow{O_iA_i}=0, \qquad \vec{v}_{O_i}+\vec{\omega}_i\times\overrightarrow{O_iB_i}=\vec{\omega}\times\overrightarrow{OB_i},
\qquad i=1,...,n,
\]
that is,
\begin{equation}\label{VEZE}
\vec{v}_{O_i}=\pm r\vec{\omega}_i\times\vec{\gamma}_i, \qquad \vec{v}_{O_i}=(R\pm 2r)\vec{\omega}\times\vec{\gamma}_i \pm r\vec{\gamma}_i  \times \vec{\omega}_i
\qquad \text{(cases I and II)}
\end{equation}
 and
\begin{equation}\label{VEZEIII}
\vec{v}_{O_1}=\pm r\vec{\omega}_1\times\vec{\gamma}_1, \qquad \vec{v}_{O_1}=\pm(2r-R)\vec{\omega}\times\vec{\gamma}_1
\pm r \vec{\gamma}_1 \times \vec{\omega}_1 \qquad
\text{(cases III and IV)}.
\end{equation}

The dimension of the configuration space $Q$ is $5n+3$. There are $4n$ independent constraints in \eqref{VEZE}, defining
a nonintegrable distribution  $\mathcal D\subset TQ$ of rank $n+3$.
The phase space of the system,
$\mathcal D$ considered as a submanifold of $TQ$,
 has the dimension $6n+6$.

Note that there are two nonholomic systems which are close to the spherical ball bearings. One is the so-called spherical support system, introduced by Fedorov in \cite{F1}. It describes the rolling without slipping of a dynamically nonsymmetric sphere $\mathbf S$ over $n$ homogeneous balls $\mathbf B_1, \dots,\mathbf B_n$ of possibly different radii, but with fixed centers. The second one is the rolling of a homogeneous ball $\mathbf B$ over a dynamically asymmetric sphere $\mathbf S$, introduced by
Borisov, Kilin, and Mamaev in \cite{BKM}.

\subsection{Symmetries}
Let $\mathbb I$ be the inertia operator of the sphere $\mathbf S$.
We choose the moving frame $O\vec{\mathbf e}_1,\vec{\mathbf e}_2,\vec{\mathbf e}_3$,
such that $O\vec{\mathbf e}_1$, $O\vec{\mathbf e}_2$, $O\vec{\mathbf e}_3$ are the principal axes of inertia: $\mathbb I=\diag(A,B,C)$.
Let $\mathbb I_i=\diag(I_i,I_i,I_i)$ and $m_i$  be the inertia operator and the mass of the $i$-th ball $\mathbf B_i$.
Then $\langle \mathbb I_i \vec{W}_i,\vec{W}_i\rangle=I_i \langle \vec{W}_i,\vec{W}_i\rangle=I_i \langle \vec{\omega}_i,\vec{\omega}_i\rangle$ and
the kinetic energy, which plays the role of the Lagrangian,   is given by:
\begin{align*}
T=\frac12 \langle \mathbb I\vec\Omega,\vec\Omega\rangle+\frac12 \sum_{i=1}^n I_i \langle \vec{\omega}_i,\vec{\omega}_i\rangle+\frac12 \sum_{i=1}^n m_i \langle \vec{v}_{O_i},\vec{v}_{O_i}\rangle.
\end{align*}

The equations of motion of the problem are given by the Lagrange-d'Alembert equations {\cite{AKN, Bloch, BMbook}}
\begin{equation}\label{L1}
\delta T=\big(\frac{\partial T}{\partial q}-\frac{d}{dt}\frac{\partial T}{\partial \dot q},\delta q\big)=0, \quad \text{for all virtual displacements} \quad \delta q\in\mathcal D_q.
\end{equation}

The kinetic energy and the constraints are invariant with respect to the $SO(3)^{n+1}$--action defined by
\begin{equation}\label{transf}
(\mathbf g,\mathbf g_1,\dots,\mathbf g_n,\vec\gamma_1,\dots,\vec\gamma_n)\longmapsto (\mathbf a\mathbf g,\mathbf a\mathbf g_1{\mathbf a}_1^{-1},\dots,
\mathbf a\mathbf g_n{\mathbf a}_n^{-1},\mathbf a\vec\gamma_1,\dots,\mathbf a\vec\gamma_n),
\end{equation}
$\mathbf a,\mathbf a_1,\dots,\mathbf a_n\in SO(3)$, representing a freedom in the choice of the reference frames
$O\vec{\mathbf e}^0_1,\vec{\mathbf e}^0_2,\vec{\mathbf e}^0_3$,
$O_i\vec{\mathbf e}^i_1,\vec{\mathbf e}^i_2,\vec{\mathbf e}^i_3$, $i=1,\dots,n$.
Also, note that \eqref{transf} does not change the vectors
\[
\vec{\Omega}_i=\mathbf g^{-1}\vec{\omega}_i, \qquad \vec{\Gamma}_i=\mathbf g^{-1}\vec{\gamma}_i, \qquad \vec{V}_{O_i}=\mathbf g^{-1}\vec{v}_{O_i}
\]
of the moving frame $O\vec{\mathbf e}_1,\vec{\mathbf e}_2,\vec{\mathbf e}_3$.

Thus, for the coordinates in the space $(TQ)/SO(3)^{n+1}$ we can take the angular velocities and the unit position vectors in the reference frame
attached to the sphere $\mathbf S$:
\[
(TQ)/SO(3)^{n+1}\cong{\mathbb R}^{3(n+1)}\times (TS^2)^n \{\vec\Omega,\vec{\Omega}_1,\dots,\vec{\Omega}_n,\dot{\vec{\Gamma}}_1,\dots,\dot{\vec{\Gamma}}_n,\vec{\Gamma}_1,\dots,\vec{\Gamma}_n\}.
\]

In the moving reference frame $O\vec{\mathbf e}_1,\vec{\mathbf e}_2,\vec{\mathbf e}_3$, the constraints become:
\begin{align}
\label{veze1}
\vec{V}_{O_i}&=(R\pm 2r)\vec{\Omega}\times\vec{\Gamma}_i \pm r\vec{\Gamma}_i \times \vec{\Omega}_i,\qquad\quad \text{(cases I and II)}\\
\label{veze1III}
\vec{V}_{O_1}&=\pm(2r-R)\vec{\Omega}\times\vec{\Gamma}_1\pm r\vec{\Gamma}_1 \times  \vec{\Omega}_1, \qquad \text{(cases III and IV)}\\
\label{veze1*}
\vec{V}_{O_i}&=\pm r\vec{\Omega}_i\times\vec{\Gamma}_i, \qquad\qquad\qquad\qquad\qquad i=1,\dots,n,
\end{align}
defining the \emph{reduced phase space} $\mathcal M=\mathcal D/SO(3)^{n+1}\subset (TQ)/SO(3)^{n+1}$ of dimension $3n+3$.

Since both the kinetic energy and the constraints are invariant with respect to the $SO(3)^{n+1}$--action \eqref{transf}, the equations of motion \eqref{L1} are also
$SO(3)^{n+1}$--invariant. Thus, they induce a well defined system on the reduced phase space $\mathcal M$. 

To simplify the constraints and the equations below, we introduce parameters
\begin{align}
\label{parametri}
&\varepsilon=\frac{R}{2R \pm 2r} \qquad \text{and} \qquad \delta=\pm\frac{R \pm 2r}{2r} \qquad \text{(cases I and II)},\\
\label{parametriIII}
&\varepsilon=\frac{R}{2R-2r} \qquad \text{and} \qquad \delta=\frac{2r-R}{2r} \qquad \quad \text{(cases III and IV)}.
\end{align}

In particular, the constraints \eqref{veze1}, \eqref{veze1III}, \eqref{veze1*} are equivalent to
\begin{equation}\label{veze2}
\vec{V}_{O_i}=\pm r\vec{\Omega}_i\times\vec{\Gamma}_i, \qquad  \vec{\Omega}_i\times\vec{\Gamma}_i=\delta\vec{\Omega}\times\vec{\Gamma}_i, \qquad i=1,\dots,n.
\end{equation}

\subsection{Equations of motion}\label{sec3}

The statements below are derived in \cite{DGJRCD} for the configuration I. The inclusion of configurations II, III, and IV can be obtain similarly and we will omit the
proofs.

Let $\vec{\mathbf F}_{B_i}$ and $\vec{\mathbf F}_{A_i}$ be the reaction forces that act on the ball $\mathbf B_i$ at the points $B_i$ and $A_i$, respectively.
The reaction force at the point $B_i$ on the sphere $\mathbf S$ is then $-\vec{\mathbf F}_{B_i}$.
By using the laws of change of angular momentum and  momentum of a rigid body in the moving reference frame for the balls $\mathbf B_i$ and the sphere $\mathbf S$ we get:
\begin{align}
\label{jednacine1}  I_i\dot{\vec{\Omega}}_i &=I_i\vec{\Omega}_i\times \vec{\Omega} \pm r\vec{\Gamma}_i\times(\vec{\mathbf F}_{B_i}-\vec{\mathbf F}_{A_i}),\\
\label{jednacine2} m_i\dot{\vec{V}}_{O_i} &=m_i\vec{V}_{O_i} \times \vec{\Omega}+\vec{\mathbf F}_{B_i}+\vec{\mathbf F}_{A_i}, \qquad\qquad\qquad i=1,...,n\\
\label{jednacine3} \mathbb I\dot{\vec{\Omega}}&=\mathbb I\vec{\Omega}\times \vec{\Omega} \mp 2r\sum_{i=1}^{n} \delta \, \vec{\Gamma}_i\times\vec{\mathbf F}_{B_i}.
\end{align}

On the other hand, from the constraint, we obtain the following kinematic equations for the unit position vectors ${\vec{\Gamma}}_i$:
\begin{equation}\label{game1}
\dot{\vec{\Gamma}}_i=\varepsilon\vec{\Gamma}_i\times\vec{\Omega}, \qquad i=1,\dots,n.
\end{equation}

As a direct consequence of the equations \eqref{jednacine1} and \eqref{game1}, we get

\begin{prop}\label{prva} The following functions are the first integrals of the equations of motion
\eqref{jednacine1}, \eqref{jednacine3}, and \eqref{game1}:
\begin{align}
\label{nova1} \langle\vec{\Gamma}_i,\vec{\Gamma}_j\rangle &=\gamma_{ij}=const, \qquad 1 \le i< j\le n,\\
\label{nova2} \langle\vec{\Omega}_i,\vec{\Gamma}_i\rangle &=c_i=const,\quad    \qquad i=1,...,n.
\end{align}
\end{prop}

The equations \eqref{nova1} are consequence of the kinematic equations \eqref{game1} only and they imply that the centers $O_i$ of the homogeneous balls $\mathbf B_i$ are in rest with respect to each other.
After fixing the values of $\gamma_{ij}$, we can consider the  relations $\langle \vec\gamma_i,\vec\gamma_j\rangle=\langle \vec\Gamma_i,\vec\Gamma_j\rangle=\gamma_{ij}$ as holonomic constraints that have no influence on the motion of the system.

From  \eqref{nova2}, we also get that the reduced phase space $\mathcal M=\mathcal D/SO(3)^{n+1}$ is foliated on $2n+3$--dimensional invariant varieties
\[
\mathcal M_c: \qquad \langle\vec{\Omega}_i,\vec{\Gamma}_i\rangle=c_i=const, \qquad i=1,...,n.
\]

By using the constraints, the vector-functions
$\vec{\Omega}_i$ can be uniquely expressed as functions of $\vec{\Omega}$, $\vec{\Gamma}_i$ on the invariant variety $\mathcal M_c$:
\begin{equation}\label{omegai1}
\vec{\Omega}_i=c_i\vec{\Gamma}_i+\delta\vec{\Omega}-\delta\langle\vec{\Gamma}_i,\vec{\Omega}\rangle\vec{\Gamma}_i.
\end{equation}

Whence, $\vec\Omega$ determines all velocities of the system on $\mathcal M_c$ and $\mathcal M_c$ is diffeomorphic
to the \emph{second reduced phase space}
\[
\mathcal N=\R^3\times \big(S^2\big)^{n}\{\Omega,\vec\Gamma_1,\dots,\vec\Gamma_n\}.
\]

This can be seen as follows. Consider the natural projection
\begin{align*}
&\pi\colon (TQ)/SO(3)^{n+1}\cong{\mathbb R}^{3(n+1)}\times (TS^2)^n \to \mathcal N,\\
&\pi(\vec\Omega,\vec{\Omega}_1,\dots,\vec{\Omega}_n,\dot{\vec{\Gamma}}_1,\dots,\dot{\vec{\Gamma}}_n,\vec{\Gamma}_1,\dots,\vec{\Gamma}_n)=(\vec\Omega,\vec{\Gamma}_1,\dots,\vec{\Gamma}_n),
\end{align*}
and let $\pi_c$ be the restriction to
$\mathcal M_c \subset \mathcal M \subset (TQ)/SO(3)^{n+1}$ of $\pi$.
Then the projection
\[
\pi_c\colon \mathcal M_c\longmapsto \mathcal N
\]
is a bijection.
Further, since $\langle\vec{\omega}_i,\vec{\gamma}_i\rangle=\langle\vec{\Omega}_i,\vec{\Gamma}_i\rangle$, we have that $\mathcal D$ is foliated
on invariant varieties
\[
\mathcal D_c\colon \qquad \langle\vec{\omega}_i,\vec{\gamma}_i\rangle=c_i, \qquad i=1,\dots,n, \qquad \dim\mathcal D_c=5n+6
\]
and $\mathcal M_c=\mathcal D_c/SO(3)^{n+1}$.
As a result we obtain the following diagram
\begin{equation*}
\label{principal}
\xymatrix@R28pt@C28pt{
\mathcal D_c \,\,\ar@{->}[d]^{/SO(3)^{n+1}} \ar@{^{(}->}[r]& \,\, \mathcal D\ar@{^{}->}[d]^{/SO(3)^{n+1}} \,\, \ar@{^{(}->}[r] & TQ=(TSO(3))^{n+1}\times (TS^2)^n \ar@{^{}->}[d]^{/SO(3)^{n+1}}\\
\mathcal M_c \,\,\ar@{->}[rrd]^{\pi_c}_{\cong} \ar@{^{(}->}[r]& \,\,\mathcal M\,\,  \ar@{^{(}->}[r] & (TQ)/SO(3)^{n+1}\cong{\mathbb R}^{3(n+1)}\times (TS^2)^n \ar@{^{}->}[d]^{\pi}  \\
 &  & \mathcal N=\R^3\times \big(S^2\big)^{n}  }
\end{equation*}
which implies that $\mathcal D_c$ and $\R^3\times SO(3)^{n+1}\times\big(S^2\big)^{n}\{\Omega,\mathbf g,\mathbf g_1,\dots,\mathbf g_n,\vec\Gamma_1,\dots,\vec\Gamma_n\}$ are diffeomorphic:
\[
\mathcal D_c\cong \R^3\times SO(3)^{n+1}\times\big(S^2\big)^{n}\{\Omega,\mathbf g,\mathbf g_1,\dots,\mathbf g_n,\vec\Gamma_1,\dots,\vec\Gamma_n\}.
\]

We define the \emph{modified operator of inertia} $\mathbf I$ as
\begin{equation}\label{modI}
\begin{aligned}
\mathbf I=\mathbb I+\delta^2\sum_{i=1}^n(I_i+m_ir^2)\pr_i,
\end{aligned}
\end{equation}
where $\pr_i\colon \R^3 \to \vec{\Gamma}_i^\perp$ is the orthogonal projection to the plane orthogonal to $\vec{\Gamma}_i$,
and set
\begin{align}
\label{m1}\vec{M}=&\, \mathbf I\vec\Omega=
\mathbb I\vec\Omega+\delta^2\sum_{i=1}^n(I_i+m_ir^2)\vec{\Omega}-\delta^2\sum_{i=1}^n(I_i+m_ir^2)\langle\vec{\Gamma}_i,\vec{\Omega}\rangle\vec{\Gamma}_i,\\
\label{n1}\vec{N}=&\, \delta\sum_{i=1}^n I_ic_i\vec{\Gamma}_i.
\end{align}

\begin{thm}\label{Glavna}
(i) The complete equations of  motion of the sphere $\mathbf S$ and the balls $\mathbf B_1,\dots,\mathbf B_n$ of the spherical ball bearings problem
on the invariant manifold $\mathcal D_c$ are given by
\begin{align}
\label{red1}\dot{\vec{M}}&=\vec{M}\times\vec{\Omega}+(1-\varepsilon)\vec N\times\vec{\Omega},\\
\label{red2}\dot{\vec\Gamma}_i&=\varepsilon\vec\Gamma_i\times\vec\Omega, \qquad\qquad\qquad i=1,\dots,n,\\
\dot{\mathbf g}&=\mathbf g \Omega,\\
\dot{\mathbf g}_i&=\mathbf g\Omega_i(\vec\Omega,\vec\Gamma_i,c_i)\mathbf g_i,
\end{align}
where $\vec M$, $\vec N$, are given by \eqref{m1} and \eqref{n1}. Here $\Omega$ and $\Omega_i(\vec\Omega,\vec\Gamma_i,c_i)$ are skew-symmetric matrices related to $\vec\Omega$ and $\vec{\Omega}_i$ after the identification \eqref{izomorfizam}; $\vec{\Omega}_i=\vec\Omega_i(\vec\Omega,\vec\Gamma_i,c_i)$ as in the equation \eqref{omegai1}.

(ii) The reduced system on $\mathcal M_c\cong \mathcal N$ is described
by the closed system \eqref{red1}, \eqref{red2}.
\end{thm}

\begin{remark}
If we formally set $\varepsilon=1$ in the system \eqref{red1}, \eqref{red2} we obtain the equation of the spherical support system introduced by
Fedorov in \cite{F1}. The system describes the rolling without slipping of a dynamically nonsymmetric sphere $\mathbf S$ over $n$ homogeneous balls $\mathbf B_1, \dots,\mathbf B_n$ of possibly different radii, but with fixed centers. It is an example of a class of
nonhamiltonian L+R systems on Lie groups with an invariant measure (see \cite{FeRCD, FJ1, Jo1}).
On the other hand, if we set $\vec N=0$, we obtain an example $\varepsilon$--modified L+R system studied in \cite{Jo2015}.
\end{remark}

\section{Invariant measure and integrable cases}\label{sec3}

\subsection{Invariant measure}
The modified inertia operator $\mathbf I$ \eqref{modI} can be rewritten as:
\[
\mathbf I=\mathbb I-\Pi, \qquad \Pi=\delta^2\sum_{i=1}^n(I_i+m_ir^2)\big(\vec{\Gamma}_i\otimes\vec{\Gamma}_i-\mathbf E\big).
\]

Along the flow of the system, $\Pi$ satisfies the matrix equation
\begin{equation}\label{GAMA}
\frac{d}{dt}\Pi=\varepsilon[\Pi,\Omega],
\end{equation}
where $\Omega$ is the skew-symmetric matrix that corresponds to the angular velocity $\vec\Omega$ via isomorphism \eqref{izomorfizam}.

\begin{thm}\label{mera}
For arbitrary values of parameters $c_i$, the reduced system \eqref{red1}, \eqref{red2} has the invariant measure
\begin{equation}\label{MERA}
\mu(\vec\Gamma_1,\dots,\vec\Gamma_n)d\Omega\wedge \sigma_1\wedge \dots \wedge \sigma_n, \qquad \mu=\sqrt{\det(\mathbf I)}=\sqrt{\det(\mathbb I-\Pi)},
\end{equation}
where $d\Omega$ and $\sigma_i$ are the standard measures on $\R^3\{\vec\Omega\}$ and $S^2\{\vec\Gamma_i\}$, $i=1,\dots,n$.
\end{thm}

Note that the existence of an invariant measure for nonholomic problems is well studied in many classical problems \cite{BM2002, BMB2013}. After Kozlov's theorem on obstruction to the existence of an
invariant measure for the variant of the classical Suslov problem (e.g., see \cite{BT2020, FJ1}) on Lie algebras \cite{Kozlov},  general existence statements
for nonholonomic systems with  symmetries are obtained in  \cite{ZenkovBloch} and \cite{FGM2015}.

A closely related problem is the integrability of nonholonomic systems \cite{AKN}.

Note that the kinetic energy of the system takes the form
\[
T=\frac12 \langle \vec M,\vec\Omega\rangle+\frac12\sum_{i=1}^n I_i c_i^2.
\]
Also, since
\[
\frac{d}{dt}\vec N=\varepsilon \vec N\times \vec\Omega,
\]
the equation \eqref{red1} is equivalent to
\begin{equation}
\frac{d}{dt}(\vec M+\vec N)=(\vec M+\vec N)\times\vec\Omega.
\label{red1*}
\end{equation}

From the above considerations we get:

\begin{prop}\label{integrali}
The system \eqref{red1}, \eqref{red2} always has the following first integrals
\[
F_1=\frac12 \langle \vec M,\vec\Omega\rangle, \quad F_2=\langle \vec M+\vec N, \vec M+\vec N\rangle, \quad
F_{ij}=\langle \vec\Gamma_i, \vec\Gamma_j\rangle, \quad 1\le i < j\le n.
\]
\end{prop}

Thus, in the special case $n=1$, we have the 5-dimensional phase space $\mathcal N=\R^3\times S^2\{\vec\Omega,\vec\Gamma_1\}$, and the system has
two first integrals $F_1$, $F_2$ and an invariant measure. For the integrability,  one needs to find a third independent first integral.

\subsection{System with one homogeneous ball}
We proceed with the case $n=1$.
To simplify notation, we denote $\vec{\Gamma}_1$ by $\vec\Gamma$ and set
\begin{align*}
&D=\delta^2(I_1+m_1r^2), \qquad d=\delta I_1 c_1, \qquad L=\langle \vec{\Omega},\vec{\Gamma}\rangle,\\
&\vec{\mathbf M}=\vec{M}+\vec{N}=\vec{M}+d\vec{\Gamma}=I\vec{\Omega}+D\vec{\Omega}+(d-DL)\vec\Gamma.
\end{align*}

The system \eqref{red1*}, \eqref{red2}, the operator $\mathbf I$, and its determinant now read
\begin{equation}\label{redukovane}
\dot{\vec{\mathbf M}}=\vec{\mathbf M}\times\vec\Omega, \qquad \dot{\vec{\Gamma}}=\varepsilon\vec\Gamma\times\vec\Omega,
\end{equation}
\[
\mathbf I=\mathbb I+D\mathbf E-D\vec\Gamma\otimes\vec\Gamma=\begin{pmatrix}
A+D-D\Gamma_1^2 & -D\Gamma_1\Gamma_2 &  -D\Gamma_1\Gamma_3\\
-D\Gamma_1\Gamma_2 & B+D-D\Gamma_2^2 & -D\Gamma_2\Gamma_3\\
 -D\Gamma_1\Gamma_3 &  -D\Gamma_2\Gamma_3 &C+D-D\Gamma_3^2
\end{pmatrix},
\]
\begin{align*}
\det(\mathbf I)&=(A+D)(B+D)(C+D)
 \big(1-D\big(\frac{\Gamma_1^2}{A+D}+\frac{\Gamma_2^2}{B+D}+\frac{\Gamma_3^2}{C+D}\big)\big).
\end{align*}

From \eqref{red1} we get
\begin{align*}
\mathbf I\dot{\vec{\Omega}}=&(\mathbb I\vec\Omega+D\vec\Omega-DL\vec\Gamma)\times\vec\Omega+(1-\varepsilon)d\vec\Gamma\times\vec\Omega+D\frac{d}{dt}\big(\vec\Gamma\otimes\vec\Gamma\big)\vec\Omega\\
  =&(\mathbb I\vec\Omega-DL\vec\Gamma)\times\vec\Omega+(1-\varepsilon)d\vec\Gamma\times\vec\Omega+\varepsilon DL \vec\Gamma\times\vec\Omega\\
  =&\mathbb I\vec\Omega\times\vec\Omega+(\varepsilon-1)(DL-d)\vec\Gamma\times\vec\Omega,
\end{align*}
implying the explicit from of the equations \eqref{redukovane}
\begin{equation}
\label{redukovane2}
\dot{\vec{\Omega}}=\mathbf I^{-1}\big(I\vec\Omega\times\vec\Omega+(\varepsilon-1)(DL-d)\vec\Gamma\times\vec\Omega\big), \qquad \dot{\vec{\Gamma}}=\varepsilon\vec\Gamma\times\vec\Omega.
\end{equation}

Thus, according to Theorem \ref{mera}, the flow of \eqref{redukovane2} preserves the measure $\sqrt{\det(\mathbf I)}d\Omega\wedge d\sigma$ on $\R^3\times S^2\{\Omega,\Gamma\}$.

For $d=0$, the equations coincide with the equations of a Chaplygin ball with inertia tensor $\mathbb I$ on a sphere ($\varepsilon\ne 1$) and the plane ($\varepsilon=1$) with slightly different definitions of parameters $\varepsilon$ and $D$.
The equations have a similar structure as the Euler-Poisson equations of Euler case of the rigid body motion about a fixed point.

\subsection{The first integrable case (generic $\mathbb I$, $\varepsilon=-1$)}
It is well known that the rolling of a Chaplygin ball over a plane ($\varepsilon=1$), for an arbitrary inertia operator $\mathbb I$, has the third integral $\langle \vec{\mathbf M},\vec{\Gamma}\rangle$. That is why, for $\varepsilon\ne 1$ we are looking for the integral of the form
\[
F_3=x_1 \mathbf M_1\Gamma_1+ x_2\mathbf M_2 \Gamma_2+x_3\mathbf M_3\Gamma_3.
\]

Along the flow of the system \eqref{redukovane} we have
\begin{equation}\label{dotF3}
\begin{aligned}
\dot{F}_3=&\Gamma_1\Omega_2\Omega_3\big((B-C)x_1-\varepsilon(B+D)x_2+\varepsilon(C+D)x_3\big) \\
&+\Omega_1\Gamma_2\Omega_3\big(\varepsilon(A+D) x_1+(C-A)x_2-\varepsilon(C+D)x_3\big)\\
&+\Omega_1\Omega_2\Gamma_3\big(-\varepsilon(A+D)x_1+\varepsilon(B+D)x_2+(A-B)x_3\big)\\
&+\Gamma_1\Gamma_2\Omega_3\big((d+\varepsilon d)x_1+(-d-\varepsilon d)x_2\big)\\
&+\Gamma_1\Omega_2\Gamma_3\big((-d-\varepsilon d)x_1+(d+\varepsilon d)x_3\big)\\
&+\Omega_1\Gamma_2\Gamma_3\big((d+\varepsilon d)x_2+(-d-\varepsilon d) x_3 \big)\\
&+L\Gamma_1\Gamma_2\Omega_3\big((-D-\varepsilon D)x_1+(D+\varepsilon D)x_2 \big)\\
&+L\Gamma_1\Omega_2\Gamma_3\big((D+\varepsilon D)x_1+ (-D-\varepsilon D)x_3\big)\\
&+L\Omega_1\Gamma_2\Gamma_3\big((-D-\varepsilon D)x_2+(D+\varepsilon D)x_3\big).
\end{aligned}
\end{equation}

Therefore, $\dot F_3=0$ if and only if the parameters $x_1,x_2,x_3$ satisfy the system of 9 homogeneous linear equations corresponding to the 9 terms given above.
Since $D\ne 0$, if $\varepsilon\ne-1$, from the last three equations we get $x_1=x_2=x_3=x$. We can take $x=1$.
The 4th, 5th and the 6th equation is then also satisfied, while from the
first 3 equations we obtain the conditions on the parameters $A, B, C$:
\begin{align*}
& (1-\varepsilon) B+ (\varepsilon -1)C=0 \\
& (\varepsilon-1) A+(1-\varepsilon)C=0   \\
& (1-\varepsilon) A+ (\varepsilon -1)B=0.
\end{align*}
Thus,  if $\varepsilon=1$, the function $\langle\vec{\mathbf M},\vec{\Gamma}\rangle$ is the integral of the equation
\eqref{redukovane} (for any  $d\in\R$), while for $\varepsilon\ne 1$, we get that $F_3=\langle\vec{\mathbf M},\vec{\Gamma}\rangle$ is the integral in the totally symmetric case $A=B=C$.
However, then $F_1$, $F_2$, $F_3$ are functionally dependent.

On the other hand, for $\varepsilon=-1$, the last 6 equations become trivial, while the first 3 have a nontrivial solution
\[
x_1=B+C-A+D, \qquad x_2=A+C-B+D, \qquad x_3=A+B-C+D.
\]

As a result, since $\varepsilon=-1$ in the  configuration III for $2r=3R$,
we get the following statement.

\begin{thm}
The spherical ball bearings problem \eqref{redukovane} in the configuration III, when $2r=3R$, i.e., the radius of the moving sphere $\mathbf S$ is twice the radius of the fixed sphere $\mathbf S_0$, is integrable. The third integral is
\[
F_3=(B+C-A+D) \mathbf M_1\Gamma_1 +(A+C-B+D) \mathbf M_2 \Gamma_2+(A+B-C+D)\mathbf M_3\Gamma_3.
\]
\end{thm}

Thus, for $d=0$, the integral $F_3$ reduces to the one found by Borisov and Fedorov for the rolling of a Chaplygin ball over a sphere \cite{BF}.

\subsection{The second integrable case ($B=C$, generic $\varepsilon$)}
Further, note that for $B=C$, the density of the measure becomes the function of $\Gamma_1$ only:
\begin{align*}
\rho=\rho(\Gamma_1)=&\frac{\sqrt{\det(\mathbf I)}}{\sqrt{C+D}}=\sqrt{(A+D)(C+D)\big(1-\frac{D\Gamma_1^2}{A+D}-\frac{D-D\Gamma_1^2}{C+D}\big)}\\
=&\sqrt{C(A+D)+D(A-C)\Gamma_1^2}.
\end{align*}

Also, for the motion of a symmetric Chaplygin ball ($B=C$) over a plane, we have an integral of the form
(up to multiplication by a constant, see \cite{DGJjpa, Kuleshov, BM2002})
\[
f=\rho^2(\Gamma_1)\Omega^2_1=C(A+D)\Omega_1^2+D(A-C)\Omega_1^2\Gamma_1^2.
\]

It appears that in the study of the rolling of the symmetric Chaplygin ball over a sphere, it is convenient to use variables $F$ and $G$ defined by
(see \cite{BM2002})
\begin{align*}
F=&\rho(\Gamma_1)\Omega_1, \qquad (f=F^2) \\
G=& A\Omega_1\Gamma_1+C(\Omega_2\Gamma_2+\Omega_3\Gamma_3)=(A-C)\Omega_1\Gamma_1+CL \qquad \big(L=\frac{G+(C-A)\Omega_1\Gamma_1}{C}\big).
\end{align*}

We are going to determine the time derivatives of $F$ and $G$ along the flow \eqref{redukovane}, i.e., \eqref{redukovane2}.

Note that $G=\langle\vec{\mathbf M},\vec{\Gamma}\rangle-d$.
Therefore, from the equation \eqref{dotF3}, where we set $x_1=x_2=x_3=1$, $B=C$, we get
\begin{align*}
\dot G=&(\varepsilon-1)(A-C)\Omega_1\Gamma_2\Omega_3-(\varepsilon-1)(A-C)\Omega_1\Omega_2\Gamma_3\\
=&(\varepsilon-1)(A-C)\Omega_1\big(\Gamma_2\Omega_3-\Omega_2\Gamma_3\big)\\
=&\frac{\varepsilon-1}{\varepsilon}(A-C)\Omega_1\dot\Gamma_1=(\varepsilon-1)(A-C)F\frac{\dot\Gamma_1}{\varepsilon \rho}.
\end{align*}

In order to find $\dot F$ we need to use equations \eqref{redukovane2}. We have
\begin{align*}
\begin{pmatrix}
\dot\Omega_1\\
\dot\Omega_2\\
\dot\Omega_3
\end{pmatrix}
=\frac{1}{\det(\mathbf I)} &
\begin{pmatrix}
\Delta_1 & (C+D)D\Gamma_1\Gamma_2 &  (C+D)D\Gamma_1\Gamma_3\\
(C+D)D\Gamma_1\Gamma_2 & \Delta_2 &  (A+D)D\Gamma_2\Gamma_3\\
(C+D)D\Gamma_1\Gamma_3 &  (A+D)D\Gamma_2\Gamma_3 &\Delta_3
\end{pmatrix}
 \cdot \\
& \cdot \begin{pmatrix}
(\varepsilon-1)(DL-d)(\Gamma_2\Omega_3-\Gamma_3\Omega_2) \\
(\varepsilon-1)(DL-d)(\Gamma_3\Omega_1-\Gamma_1\Omega_3)+(C-A)\Omega_1\Omega_3   \\
(\varepsilon-1)(DL-d)(\Gamma_1\Omega_2-\Gamma_2\Omega_1)+(A-C)\Omega_1\Omega_2
\end{pmatrix},
\end{align*}
where
\begin{align*}
&\Delta_1=(C+D)^2-(C+D)D(\Gamma_2^2+\Gamma_3^2),\\
&\Delta_2=(A+D)(C+D)-(C+D)D\Gamma_1^2-(A+D)D\Gamma_3^2,\\
&\Delta_3=(A+D)(C+D)-(C+D)D\Gamma_1^2-(A+D)D\Gamma_2^2.
\end{align*}

Therefore
\begin{align*}
\dot\Omega_1=&\frac{1}{\rho^2}((C+D)-D(1-\Gamma_1^2))(\varepsilon-1)(DL-d)(\Gamma_2\Omega_3-\Gamma_3\Omega_2)\\
&+\frac{1}{\rho^2}D\Gamma_1\Gamma_2((\varepsilon-1)(DL-d)(\Gamma_3\Omega_1-\Gamma_1\Omega_3)+(C-A)\Omega_1\Omega_3)\\
&+\frac{1}{\rho^2}D\Gamma_1\Gamma_3((\varepsilon-1)(DL-d)(\Gamma_1\Omega_2-\Gamma_2\Omega_1)+(A-C)\Omega_1\Omega_2)\\
=&\frac{1}{\rho^2}\big(C(\varepsilon-1)(DL-d)(\Gamma_2\Omega_3-\Gamma_3\Omega_2)+D\Gamma_1\Omega_1(C-A)(\Gamma_2\Omega_3-\Gamma_3\Omega_2)\big)\\
=&\frac{1}{\varepsilon\rho^2}\big(C(\varepsilon-1)(DL-d)+D\Gamma_1\Omega_1(C-A)\big)\dot\Gamma_1\\
=& \frac{1}{\varepsilon\rho^2}\big(D(\varepsilon-1)(G+(C-A)\Omega_1\Gamma_1)+D\Gamma_1\Omega_1(C-A)-Cd(\varepsilon-1)\big)\dot\Gamma_1\\
=& \frac{1}{\varepsilon\rho^2}\big(D(\varepsilon-1)G+\varepsilon D\Gamma_1\Omega_1(C-A)-Cd(\varepsilon-1)\big)\dot\Gamma_1
\end{align*}
and
\begin{align*}
\dot F=&\dot\rho\Omega_1+\rho\dot\Omega_1\\
=&\frac{1}{\rho}D(A-C)\Gamma_1\dot\Gamma_1\Omega_1+\frac{1}{\varepsilon\rho}\big(D(\varepsilon-1)G+\varepsilon D\Gamma_1\Omega_1(C-A)-Cd(\varepsilon-1)\big)\dot\Gamma_1\\
=&\big(D(\varepsilon-1)G-Cd(\varepsilon-1)\big)\frac{\dot\Gamma_1}{\varepsilon\rho}
\end{align*}

Let $\Phi(\Gamma_1)$ be a primitive function of $1/{\varepsilon\rho}$:
\[
\frac{d\Phi}{d\Gamma_1}=\frac{1}{\varepsilon\rho} \qquad \big(\dot\Phi=\frac{\dot\Gamma_1}{\varepsilon\rho}\big).
\]

Due to the structure of the expressions for $\dot F$ and $\dot G$, we are looking for a third first integral in the form
\[
F_3=\big(y_1 F+ y_2 G + y_3\big)\exp(y_4\Phi).
\]

We have
\begin{align*}
\dot F_3=&\big(y_1 \dot F+ y_2 \dot G\big)\exp(y_4\Phi)+y_4\big(y_1 F+ y_2 G + y_3\big)\dot\Phi\exp(y_4\Phi)\\
=&\big(y_1 \big(D(\varepsilon-1)G-Cd(\varepsilon-1)\big) + y_2 (\varepsilon-1)(A-C)F\big)\dot\Phi\exp(y_4\Phi)\\
&+y_4\big(y_1 F+ y_2 G + y_3\big)\dot\Phi\exp(y_4\Phi)
\end{align*}

Therefore, if $y_1,y_2,y_3,y_4$ are solutions of the system
\begin{align}
\label{Ieq} (\varepsilon-1)(A-C)y_2+y_1y_4=&0,\\
\label{IIeq} D(\varepsilon-1)y_1+y_2y_4=&0,\\
\label{IIIeq} -Cd(\varepsilon-1)y_1+y_3y_4=&0,
\end{align}
the function $F_3$ is a first integral of the system \eqref{redukovane} for $B=C$.

By dividing equations \eqref{Ieq} and \eqref{IIeq} we get
\[
\frac{A-C}{D}\frac{y_2}{y_1}=\frac{y_1}{y_2}.
\]
Whence
\[
y_1=\pm \sqrt{\frac{A-C}{D}} y_2 \qquad \text{and} \qquad y_4=\pm (1-\varepsilon)\sqrt{\frac{A-C}{D}} y_2.
\]

On the other hand, by dividing equations \eqref{IIeq} and \eqref{IIIeq} we obtain
\[
y_3=-y_2\frac{Cd}{D}.
\]
Here $y_2\ne 0$ is arbitrary. By taking $y_2=D$ we get

\begin{thm}
The spherical ball bearings problem \eqref{redukovane} for $B=C$ is integrable for all $\varepsilon$.  Along with $F_1$ and $F_2$, the system
has  two additional, nonalgebraic first integrals $F_3$ and $F_4$:
\begin{equation}\label{f3pm}
F_{3,4} =\big(\pm\sqrt{D(A-C)} F+ D G -dC \big)\exp(\pm(1-\varepsilon)\sqrt{D(A-C)}\Phi).
\end{equation}

\end{thm}

 Note that the product of the two nonalgebraic first integrals
\begin{align*}
F_3F_4=&(D G -dC)^2-D(A-C)F^2\\
          =& D^2G^2-2dCDG+d^2C^2-D(A-C)\big(C(A+D)\Omega_1^2+D(A-C)\Omega_1^2\Gamma_1^2\big)
\end{align*}
is an affine combination of $F_1$ and $F_2$:
\[
F_3F_4=CD(C+D)\langle \vec{M},\vec\Omega\rangle-CD\langle\vec{\mathbf M},\vec{\mathbf M}\rangle -C(C+D)d^2.
\]

\section{Integration of the planar ball bearings}

In \cite{DGJRCD}, we also introduced  the planar ball bearings problem. It  was proved there that this system is integrable in quadratures. Here, we are going to present the procedure of its integration.
 In \cite{DGJRCD} the equations of motion were  derived for the case of three homogeneous balls. Nevertheless, it was indicated in \cite{DGJRCD} that all considerations could be adopted for the general case of $n$ balls. Here we consider this general case of $n$ balls.

The planar  $n$ balls bearing problem is a system that consists of $n$ homogeneous balls  $\mathbf B_1,\dots,\mathbf B_n$ of the radius $r$,
that roll without slipping over the fixed plane $\Sigma_0$. In addition it is assumed that the moving plane $\Sigma$
of the mass $m$ is placed over the balls, such that there is no slipping between the balls and moving plane.
Let  $O_0$ be a fixed point of the plane $\Sigma_0$  and $O, O_1,\dots, O_n$ be the centers of  masses of the plane $\Sigma$ and the balls $\mathbf B_1,\dots,\mathbf B_n$, respectively.
In the fixed reference frame, the positions of the points $O$, $O_i$ are given by
\[
O(x,y,2r), \qquad O_i(x_i,y_i,r), \qquad i=1,\dots,n.
\]

\begin{figure}[ht]\label{Fig3}
\includegraphics[width=115mm]{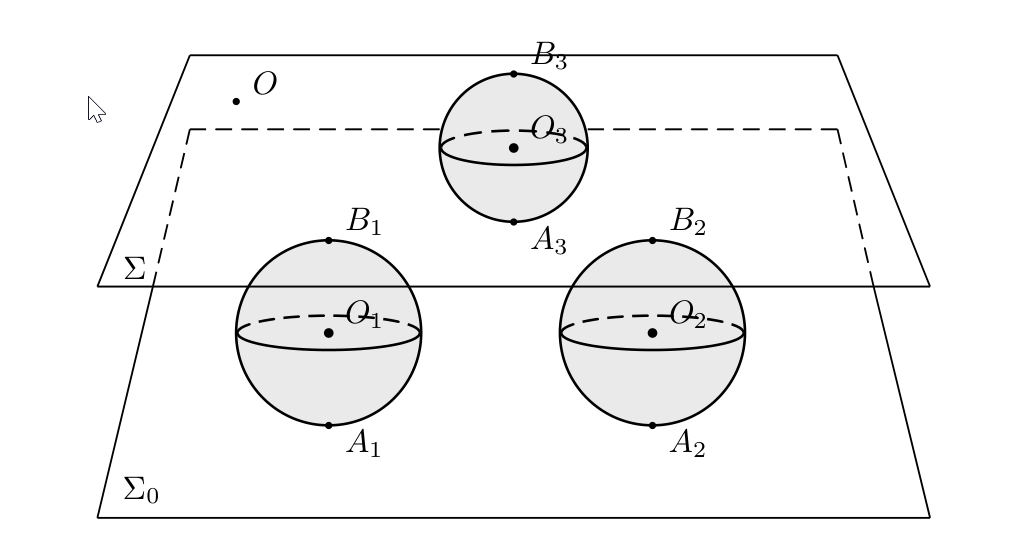}
\caption{Planar ball bearings for $n=3$}
\end{figure}

Let  $A_1,..., A_n$ and
$B_1, B_2,...,B_n$ be the contact points of the balls $\mathbf B_1,\dots,\mathbf B_n$ with the planes $\Sigma_0$ and $\Sigma$, respectively,  see Figure \ref{Fig3} for the special case $n=3$.
 It appears that the moving configuration of the centers $(O_1(t),\dots,O_n(t))$, i.e., of the contact points $(B_1(t),\dots,B_n(t))$, is congruent to
the configuration $(O_1,\dots,O_n)$ formed at the initial condition \cite{DGJRCD}. In other words, as in the spherical case,
the centers $O_i$ of the balls $\mathbf B_i$ are in rest in relation to each other along the motion.

We denote  $v_\varphi=\dot\varphi$, $v_x=\dot x$, $v_y=\dot y$
and  introduce
\begin{equation}\label{oznake}
\begin{aligned}
\vec N=&(N_1,N_2,0)=\sum_{i=1}^n \delta_i\overrightarrow{OB_i}, \qquad M=\sum_{i=1}^n \delta_i\langle\overrightarrow{OB_i},\overrightarrow{OB_i}\rangle,\\
\delta_i=&\frac{m_ir^2+I_i}{4r^2}, \qquad  i=1,\dots,n, \qquad \delta=\delta_1+\dots+\delta_n,
\end{aligned}
\end{equation}
 where $\diag(I_i,I_i,I_i)$ is the inertia operator and $m_i$ is the mass of the $i$-th ball $\mathbf B_i$.

In \cite{DGJRCD} it is proved that the equation of motion can be reduced to
\begin{equation}\label{eq:Q}
\mathcal Q=\{(v_x,v_y,v_\varphi,N_1,N_2,M)\in \R^6\,\vert\, \delta M > N_1^2+N_2^2\}.
\end{equation}

 Let $I$ be the moment of inertia of the plane $\Sigma$ with respect to the line perpendicular to $\Sigma$ at $O$.
If we introduce
\begin{align*}
&\mathbf v=(v_x,v_y,v_\varphi), \qquad
 \mathbf n=(N_1,N_2,M),\\
&\mathbf m=\frac12(N_1 v_\varphi^2-\delta v_\varphi v_y,N_2 v_\varphi^2+\delta v_\varphi v_x, v_\varphi(N_1 v_x+N_2 v_y)),\\
&\mathbb I=
\begin{pmatrix}
m+\delta & 0            & -N_2 \\
0        & m+\delta     &  N_1  \\
-N_2     & N_1          &  I+M
\end{pmatrix},\qquad
\mathbb J=-\frac 12
\begin{pmatrix}
\delta & 0 &      N_2\\
0     & \delta & -N_1\\
  2N_1 & 2N_2 & 0
\end{pmatrix},
\end{align*}
then the reduced equations of motion on $\mathcal Q$  become
\begin{equation}\label{REDtrikugle}
\dot{\mathbf v}=\mathbb I^{-1}\mathbf m, \qquad \dot{\mathbf n}=\mathbb J \mathbf v.
\end{equation}


  It  was proved in \cite{DGJRCD} that the equations \eqref{REDtrikugle} have the following first integrals
\begin{equation}\label{integraliRED}
\begin{aligned}
& f_1=(m+\delta) v_x- v_\varphi N_2, \\
& f_2=(m+\delta) v_y + v_\varphi N_1, \\
& f_3=\delta M-(N_1^2+N_2^2),\\
& f_4=T=\frac12 (I+M) v_\varphi^2+ \frac12 (m+\delta)(v_x^2+v_y^2)+v_\varphi(N_1 v_y-N_2 v_x).
\end{aligned}
\end{equation}
Also it was  shown in \cite{DGJRCD} that the equations \eqref{REDtrikugle} possess the invariant measure
\[
\sqrt{\det(\mathbb I)}\, dv_x \wedge dv_y \wedge dv_\varphi \wedge dN_1 \wedge dN_2 \wedge dM,
\]
where
\[
\det (\mathbb I)=(m+\delta)\big((m+\delta)I + m M+ (\delta M-(N_1^2+N_2^2) \big)> 0\vert_\mathcal Q.
\]
The system \eqref{REDtrikugle} can be solved by quadratures.

At  an invariant level set of the three first integrals
\[
\mathcal Q_d\colon \qquad f_1=d_1,\qquad f_2=d_2, \qquad f_3=d_3,
\]
where $d=(d_1, d_2, d_3)$ are given constants, we obtain a closed system in the space $\R^3\{v_\varphi,N_1,N_2\}$ given by
\begin{equation}\label{trikugle*}
\begin{aligned}
\dot{v}_\varphi=&\frac{m v_\varphi(N_1 {d_1}+N_2 {d_2})}{2\det(\mathbb I)},\\
\dot N_1
=&-\frac{m+2\delta}{2(m+\delta)}N_2 v_{\varphi}-\frac{\delta d_1}{2(m+\delta)},\\
\dot N_2
=&\frac{m+2\delta}{2(m+\delta)}N_1 v_{\varphi}-\frac{\delta d_2}{2(m+\delta)},
\end{aligned}
\end{equation}
where
\[
\det (\mathbb I)=(m+\delta)\big((m+\delta)I + \frac{m}{\delta}(N_1^2+N_2^2)+ \frac{m d_3}{\delta}+d_3 \big).
\]

In order to integrate the system \eqref{trikugle*},  we introduce the polar coordinates
\[
N_1=A\cos\theta, \qquad N_2=A\sin\theta.
\]
In  the new coordinates the equations \eqref{trikugle*}  become
\begin{equation}\label{eq:nove}
\begin{aligned}
\dot{v}_\varphi=&\frac{m v_\varphi A}{2(d_5+\frac{m(m+\delta)}{\delta}A^2)}({d_1}\cos\theta+{d_2}\sin\theta),\\
\dot A=& -\frac{\delta}{2(m+\delta)}({d_1}\cos\theta+{d_2}\sin\theta)\\
A\dot \theta=& \frac{m+2\delta}{2(m+\delta)}Av_{\varphi}+\frac{\delta}{2(m+\delta)}(d_1\sin\theta-d_2\cos\theta)
\end{aligned}
\end{equation}

From  the first two equations, when $\dot{A}\ne 0$, one gets
$$
\frac{d v_\varphi}{d A}=-\frac{m(m+\delta)}{\delta}\frac{Av_\varphi}{d_5+\frac{m(m+\delta)}{\delta}A^2}.
$$
 The integration leads to
\begin{equation}\label{vprekoA}
v_\varphi=v_\varphi(A)=\frac{d_6}{\sqrt{d_5+\frac{m(m+\delta)}{\delta}A^2}},
\end{equation}
where $d_6$ is  the constant of integration.

The case when $\dot{A}=0$ is considered in \cite{DGJRCD} (see Remark 4 in \cite{DGJRCD}).

\begin{remark}  Equation \eqref{vprekoA} gives the first integral
\[
F=\frac{m(m+\delta)}{\delta}v_\varphi^2 A^2+\frac{(I\delta+d_3)(m+\delta)^2}{\delta}v_\varphi^2=d_6=const,
\]
 a modification of the energy integral that in  the new coordinates has the form
\[
f_4=\frac{m}{2\delta(m+\delta)}v_\varphi^2A^2+\frac{I\delta+d_3}{2\delta}v_\varphi^2+\frac{d_1^2+d_2^2}{2(m+\delta)}.
\]
 Thus, we have $F=2(m+\delta)^2f_4-2(m+\delta)(d_1^2+d_2^2)$.
\end{remark}

Let us introduce the constant $\alpha$ such that
\begin{align*}
& d_1\cos\theta+d_2\sin\theta=\sqrt{d_1^2+d_2^2}\ \cos(\theta-\alpha),\\
& d_1\sin\theta-d_2\cos\theta=\sqrt{d_1^2+d_2^2}\ \sin(\theta-\alpha).
\end{align*}

By using \eqref{vprekoA}, equations \eqref{eq:nove} reduce to
\begin{equation}\label{eq:atheta}
\begin{aligned}
\dot A=& -k\cos(\theta-\alpha),\\
A\dot \theta=& k\sin(\theta-\alpha)+\frac{m+2\delta}{2(m+\delta)}Av_{\varphi}(A),
\end{aligned}
\end{equation}
where $k=\frac{\delta}{2(m+\delta)}\sqrt{d_1^2+d_2^2}$.

As we mentioned, the original system has an invariant measure with the density
$ \mu=\sqrt{\det(\mathbb I)}$.
One  can check that the density of the invariant measure for the last equations reduces to $ \mu= A$.
 We will use this observation about the invariant measure to finish the integration.

Equations \eqref{eq:atheta} can be rewritten  in the form:
\[
\frac{dA}{-k\cos(\theta-\alpha)}=\frac{A\ d\theta}{k\sin(\theta-\alpha)+\frac{m+2\delta}{2(m+\delta)}Av_{\varphi}(A)},
\]
or equivalently
\[
\big(k\sin(\theta-\alpha)+\frac{m+2\delta}{2(m+\delta)}Av_\varphi (A)\big)dA +kA\cos(\theta-\alpha)d\theta=0.
\]

 We observe that the left-hand side of the last formula presents a total differential. Thus, we finally get
\[
kA\sin(\theta-\alpha)+\frac{(m+2\delta)\delta}{2m(m+\delta)^2}\sqrt{d_5+\frac{m(m+\delta)}{\delta}A^2}=d_7=const.
\]
From the last equation  we express $\theta$ as a function of $A$. By plugging it in the system \eqref{eq:atheta} and performing one more integration, we calculate $A$ as a function of time. From
\eqref{vprekoA} we find also $v_\varphi$ as a function of time.

 Finally, we note that it would be interesting to study
the above nonholonomic systems with added gyroscopes (e.g., see \cite{Zhuk1893, DGJ, BMbook}), and also to study variations of the problems in $\mathbb R^d$ with an
arbitrary dimension $d>3$ (e.g., see \cite{FK1995, Sch, FJ1, Naranjo2019b, Naranjo2019a, Jov2019}).

\subsection*{Acknowledgements}
This research has been supported by the Project no. 7744592 MEGIC "Integrability and Extremal Problems in Mechanics, Geometry and Combinatorics" of the Science Fund of Serbia, Mathematical Institute of the Serbian Academy of Sciences and Arts and the Ministry for Education, Science, and Technological Development of Serbia, and the Simons Foundation grant no. 854861.

\end{document}